\documentstyle[pre,aps,multicol]{revtex}

\input{epsf}

\title{Aging dynamics in a\\ colloidal glass of Laponite}

\author{B\'ereng\`ere Abou, Daniel Bonn and Jacques Meunier}
\address{Laboratoire de Physique Statistique\\
Ecole Normale Sup\'erieure \\
24, rue Lhomond, 75231 Paris, FRANCE}
\begin{document}
\maketitle


\begin{abstract}

\noindent The aging dynamics of colloidal suspensions of Laponite, 
a synthetic clay, is investigated using dynamic light stattering (DLS) 
and viscometry after a quench into the glassy phase. DLS allows to
follow 
the diffusion of Laponite particles and reveals that there are two 
modes of relaxation. The fast mode corresponds to a rapid diffusion of 
particles within "cages" formed by the neighboring particles. The slow
mode 
corresponds to escape from the cages: its average relaxation 
time increases exponentially fast with the age of the glass. In
addition, 
the slow mode has a broad distribution of relaxation times, its
distribution 
becoming larger as the system ages. Measuring the concomitant increase
of 
viscosity as the system ages, we can relate the slowing down of the
particle dynamics to the viscosity.
\end{abstract}
\begin{multicols}{2}

\section{INTRODUCTION}

Glassy systems are characterized by an equilibration time that is much
longer than any laboratory time scale: glasses are out of equilibrium
systems. One of the inte\-resting consequences is that, in any case for
short times after a quench into the glassy phase, the properties of a
glass may still evolve on an observable time scale: we say that the
system {\em ages}. The typical relaxation time of the glass does not become infinite immediately, but grows with the waiting time $t_w$, i.e., the time expired after the quench into the glassy phase was done. Understanding the aging processes is of great interest as the glass transition is still ill-understood.

Recent progress in mode-coupling theory has allowed for a first detailed description of the aging process~\cite{bouchaud1996}. The aging dynamics of the system is described in term of the evolution of the correlation and response functions with the waiting time. In most systems such as structural glasses, unfortunately correlation and response functions are not easy to obtain experimentally. Testing the applicability of the new mode coupling theory has therefore been limited to specific theoretical models and simulations (Lennard-Jones glass) studies~\cite{bouchaud1996,kob1997}.

Colloidal glasses can be looked upon as model glasses compared to
structural glasses where there are two coupled control parameters,
density and temperature. In colloidal glasses, the volume fraction of the particles is one of the  control
parameters that can be varied independently from the others. Because of
the experimental difficulties in studying structural glasses,
colloidal glasses have been studied extensively. In fact, it is
probably fair to say that the strongest evidence for the applicability
of classical mode coupling theory (that applies only above the glass
transition temperature and therefore cannot describe aging) comes from
colloidal systems~\cite{nelson,pusey1987}. 

In order to see whether
colloidal glasses can also be used to investigate aging processes, we
study colloidal suspensions of Laponite, a synthetic  clay composed of
monodisperse disc-shaped particles. These colloidal suspensions form
very viscous phases at low particle
concentrations~\cite{mourchid1995,mourchid1998lang}. The first evidence
of aging in colloidal glasses of Laponite was reported in~\cite{bonn-aging}. In this paper, we present quantitative results
of aging dynamics in a glassy suspension of Laponite using dynamic
light scattering measurements and viscometry.

\section{PREPARATION AND VISCOMETRY}

The experiments are performed with aqueous suspensions of Laponite RD, a hectorite synthetic clay provided by Laporte Ltd. The particles are colloidal disks of $25$ nm diameter and $1$ nm thickness~\cite{kroon1998}, with a negative surface charge on both faces. Laponite powder was mixed with ultra-pure water at pH =10 obtained by addition of NaOH, providing chemically stable particles~\cite{thompson1992,mour1998}. The suspension was stirred vigorously during 15 minutes and then filtered through a Millipore Millex-AA 0.8 $\mu$m filter unit. This preparation procedure allows us to obtain a reproducible initial liquid state. Evaporation of water or CO$_2$ contamination of the sample was completely avoided by covering the sample with a layer of vaseline oil. The aging time $t_w=0$ is defined as the moment the suspension passes through the filter. 

Suspensions prepared in this way are optically transparent and are initially liquid. Within a time varying from a few minutes to a few hours for the different concentrations considered here ($2\%$ wt to $5\%$ wt), a three order of magnitude increase in viscosity is observed : the suspension does not flow anymore when tumbling the recipient. Since the physical properties of the dispersion depend on the time after preparation $t_w$, the conclusion is that the samples age.
That the origin of this aging behavior is due to the glassy dynamics
was shown by Bonn {\it et al.}~\cite{bonn-aging,bonn-filter}. They showed that Laponite
suspensions form glasses and not gels, as had been proposed
before~\cite{mour1998,pignon1997}. Bonn {\it et al.}~\cite{bonn-filter} studied the structure of Laponite
suspensions using static light scattering. Contrary to previous
observations, they found no evidence for a fractal-like organization of
the particles, provided the suspensions were filtered. They
consequently showed that the large viscosity increase is {\em not} due
to the formation of a fractal network. In addition, it has been shown recently that Laponite powder dissolves
quickly into both individual particles and aggregates of a few particles dispersing more slowly~\cite{nicolai2000}. At the final state of dispersion (about 5 hours), the system contains aggregates responsible for the strong scattering at small scattering vectors. In our experiments, aggregates of particles are broken up by the filtration.

This colloidal glass is obtained for very low volume fractions $\Phi
\simeq 0.01$ compared to those for usual spherical colloids, for which
glasses are obtained above $\Phi \simeq 0.5$~\cite{pusey1987}. To account for this difference, it was proposed~\cite{bonn-aging} that the Laponite suspensions form so-called Wigner glasses, very low-density glasses whose formation is due to the existence of long-range Coulombic repulsions~\cite{wigner1938,bosse1998,lai1997}. These repulsions originate from the strong surface charges at the faces of the colloidal disks, making that the effective volume fraction is very high. Recent experiments have shown that the location of the "glass transition" line in the (volume fraction / electrolyte concentration) phase diagram is consistent with this assumption~\cite{levitz2000}.

\begin{figure}
\centerline{\epsfxsize=8.6 truecm \epsfbox{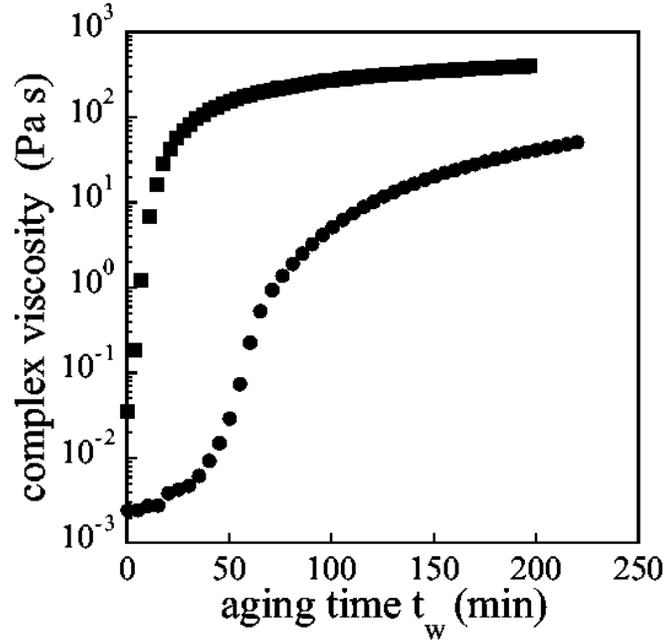}}
\caption{Complex viscosity as a function of the aging time $t_w$ for
Laponite suspensions at $2.5\%$ wt (circles) and $3.5\%$ wt
(squares). The visco-elastic moduli are measured on a Reologica
Stress-Tech rheometer at a pulsation $\omega=1$ s$^{-1}$ and with an imposed strain $\gamma_0=0.01$.}
\label{complexvis}
\end{figure}

The Laponite suspensions age on time scales that depend on the particle concentration. Oscillatory shear experiments were performed on a controlled stress (Reologica StressTech) rheometer using a Couette geometry with a 1 mm gap. The sample was exposed to a sinusoidal strain $\gamma=
\gamma_0 \sin{\omega t} $, with $\gamma_0=0.01$ and $\omega= 1$
s$^{-1}$, and allowing for measurement of the visco-elastic
response. The storage modulus $G'$ and the loss modulus $G''$ were determined in the linear visco-elastic regime. At $t_w=0$, the loss and storage moduli are roughly of the same order of
magnitude with $G '' > G '$. As the system ages, the storage modulus increases more rapidly than the loss modulus, becoming easily more than two orders of magnitude larger within a typical experimental time scale. 
Figure \ref{complexvis} shows the evolution of the complex viscosity
$\eta^{*}$, as a function of the aging time $t_w$, for suspensions
with different concentrations. The complex viscosity $\eta^{*}$ was
calculated from the visco-elastic moduli, as 
$\eta ^{*}= \sqrt{G^{'2} + G ^{''2} } / \omega$. The complex viscosity of the suspension with $2.5\%$ wt
increases by three orders of magnitude on a time scale of 100 minutes
while the same increase of viscosity arises on about 10 minutes for a suspension with $3.5\%$ wt. The aging dynamics of the suspensions can therefore be varied by changing the particle concentration.

\section{DYNAMIC LIGHT SCATTERING}

The dynamics of formation of the colloidal glass was investigated
using dynamic light scattering (DLS). The light scattering set-up is
as follows: a He-Ne laser beam ($\lambda=632.8$ nm) is focused on a
cylindrical sample in an index-matching bath of Toluene
($n=1.50$). The scattered intensity transmitted through the cell is
detected by an optical fiber coupled to an avalanche photodiode under
a scattering angle $\theta$, defined as the angle between the transmitted
and the scattered beam. The signal is then analysed by a ALV-5000
logarithmic correlator, which directly calculates the normalized
intensity autocorrelation function over nine decades in time
$g_2({\bf q},t) - 1 \equiv <I({\bf q},0)I({\bf q},t)>/<I({\bf q},0)>^2 $, where $t$ is the delay time. The modulus of the scattering wave vector, defined as $q= (4 \pi n / \lambda)
sin(\theta/2)$, with $n$ the refractive index of the suspension
($n=1.33$) was between $ 9. 10^6 \mbox{m}^{-1}< q < 3. 10^7 \mbox{m}^{-1}$, as the
scattering angle was varied between $40^{\mbox{\tiny{o}}}$ and $150^{\mbox{\tiny{o}}}$. Its
inverse, $2 \pi /q$, determines the length scale probed in a DLS experiment. 

In our experiments, the intensity correlation function was recorded
while the system is aging. In such an experiment, the acquisition time
should allow for a good average of the auto-correlation function data
without a significant aging of the system occuring during the
experiment. Because of that, experiments were performed with
suspensions of sufficiently low concentration so that the aging is
slow compared to the time necessary to obtain the correlation
functions. In the following, we present results obtained with a $2.5
\%$ wt suspension. The
acquisition time was chosen to be $60$ seconds, allowing for a good
average of the whole curve. We experimentally checked that aging was
negligible during this acquisition time for the $2.5 \%$ wt suspension
by recording the auto-correlation functions just before and just after
the measurement.
 
\begin{figure}
\centerline{\epsfxsize=8.6 truecm \epsfbox{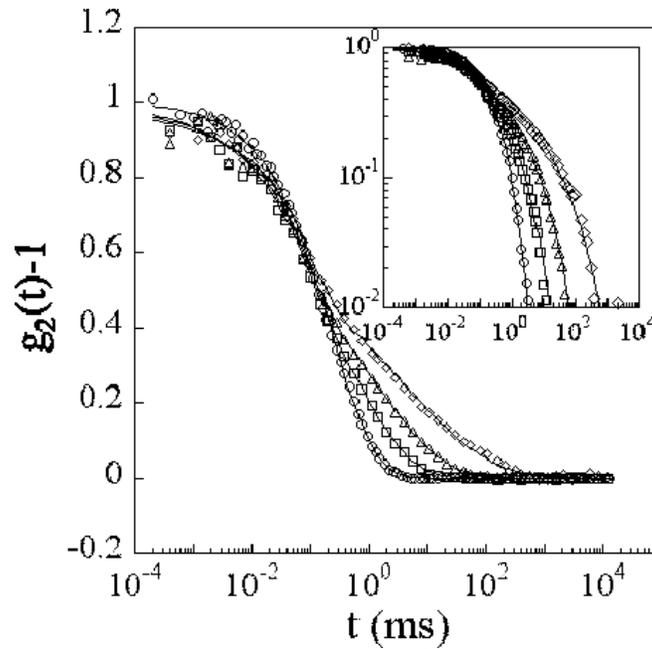}}
\caption{Autocorrelation functions of the scattered intensity by a
$2.5\%$ wt suspension and corresponding fits for aging times $t_w=0,
200, 400, 600$ min (from left to right) at a scattering angle
$\theta=90^{\mbox{\tiny{o}}}$. The autocorrelation functions and their
corresponding fits are also represented on a log-log plot in inset.}
\label{inserfit}
\end{figure}

Auto-correlation functions taken for various aging times $t_w$, under a scattering angle $\theta=90^{\mbox{\tiny{o}}}$, are shown in
Figure~\ref{inserfit}. Mainly two relaxations can be observed. The
first one, observed for short delay times $t$, is relatively fast and
appears to be independent on the aging time. The second relaxation,
observed for long delay times $t$, depends strongly on the waiting
time.
 
In order to describe the two processes quantitatively, $g_2({\bf
q},t)-1$, was fitted by a sum of an exponential and a stretched
exponential function as:
$$
g_2({\bf q},t)-1=A \exp{-(t/\tau_1)}+(1-A)\exp{-(t/\tau_2)^{\alpha}}
$$
The stretched exponential is used since it has
been found empirically that it provides a good description of the slow
relaxation processes encountered in glassy systems. The fits
corresponding to different aging times $t_w$  are shown in
Figure~\ref{inserfit} and describe the correlation functions very
well, for all aging times.

\begin{figure}
\centerline{\epsfxsize=8.6 truecm \epsfbox{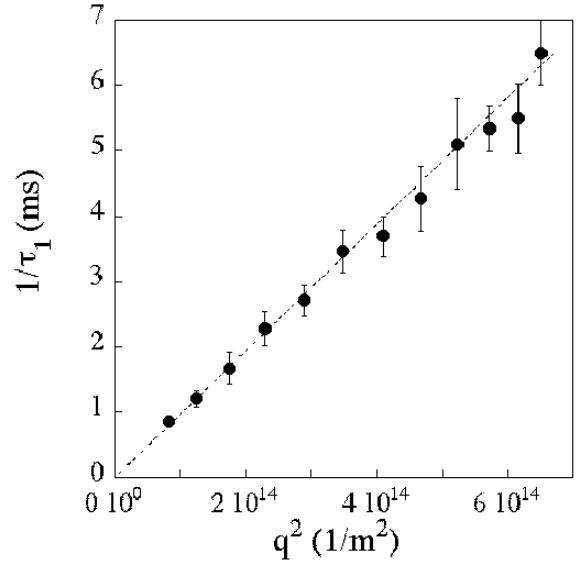}}
\caption{Inverse of the characteristic time $\tau_1$ of the fast relaxation as a
function of the modulus of the wave vector $q$, for a $2.5\%$ wt Laponite suspension.} 
\label{tau1}
\end{figure}

\begin{figure}
\centerline{\epsfxsize=8.6 truecm \epsfbox{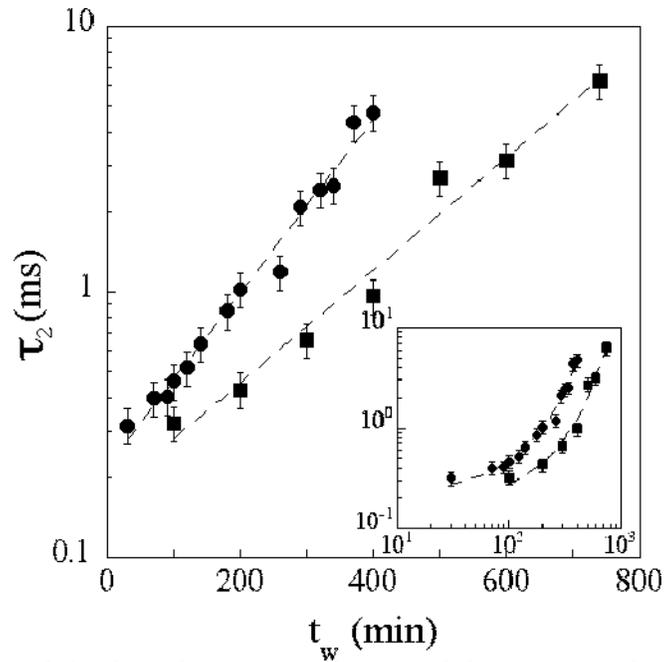}}
\caption{Characteristic time $\tau_2$ of the slow relaxation as a
function of the aging time for $2.5\%$ wt (squares) and $2.8\%$ wt
(circles) Laponite suspensions from auto-correlation functions taken under a scattering angle $\theta=90^{\mbox{\tiny{o}}}$. $\tau_2 (t_w)$ is also represented on a log-log plot in inset.}
\label{inserttau2}
\end{figure}

The first term of the fit function corresponds to a fast
relaxation. The parameter $\tau_1$ was in fact determined
independently by a linear fit of $\ln(g_2({\bf q},t)-1)$ for short times $t$,
in order to constrain the fitting procedure. This relaxation time was
found to be independent on the aging time $t_w$. Furthermore, studying
the angular dependence of the scattering, it is found that the inverse
of $\tau_1$ varies as $q^2$ as shown in Figure~\ref{tau1}. This shows
that, for short times, the Laponite particles undergo 'normal'
Brownian motion: this q-dependence is the same as that found for very
dilute colloidal dispersions~\cite{berne}; for our system, we find a collective diffusion
coefficient $D= 9.5~10^{-12}$ m$^2$ s$^{-1}$, of the same order of
magnitude as the one measured in a very dilute
solution~\cite{bonn-aging}. On the other hand, the relaxation time
$\tau_2$, that characterizes the aging, was found to increase
exponentially fast as a function of $t_w$ as $\tau_2 = \tau_0
\exp{(t_w/t_0)}$, where $\tau_0 \sim 0.1$ ms and $t_0 \sim 10^{-5}$~s (Figure~\ref{inserttau2}; see also the discussion below). As the total decay time of the correlation function is a measure for the time a particle needs to "forget" its initial position, this shows that very rapidly, the aging freezes in a certain degrees of freedom of the system. 

\begin{figure}
\centerline{\epsfxsize=8.6 truecm \epsfbox{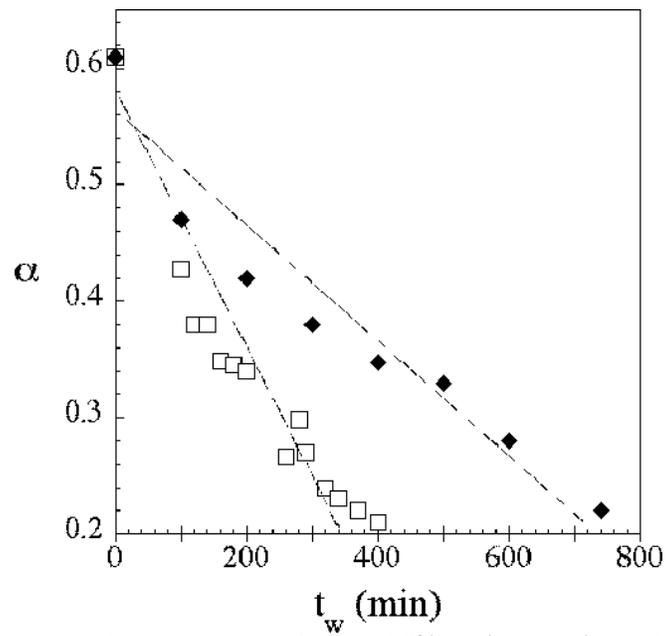}}
\caption{Stretch exponent $\alpha$ as a function of the aging time for $2.5\%$ wt (diamond) and $2.8\%$ wt (open squares) Laponite suspensions. These fit parameters correspond to auto-correlation functions taken under a scattering angle $\theta=90^{\mbox{\tiny{o}}}$.}
\label{alpha}
\end{figure}

As was the case for $\tau_1$, the relaxation time $\tau_2$ was
found also to scale with $1/ q^2$, a feature again reminiscent of
classical (Fickian) diffusion, although the stretched exponential
shows that there are strong hydrodynamic interactions between the
particles. Finally, the stretch exponent $\alpha$ was found to depend only on
the aging time and not on the scattering wavevector $q$. The exponent
$\alpha$ decreases roughly linearly between $1$ and $0$, as shown in Figure \ref{alpha}. The aging time $t_w$ for
which $\alpha \simeq 0$ corresponds to the ergodic-nonergodic
transition on the laboratory time scale~\cite{bonn-aging}. 

\begin{figure}
\centerline{\epsfxsize=7truecm \epsfbox{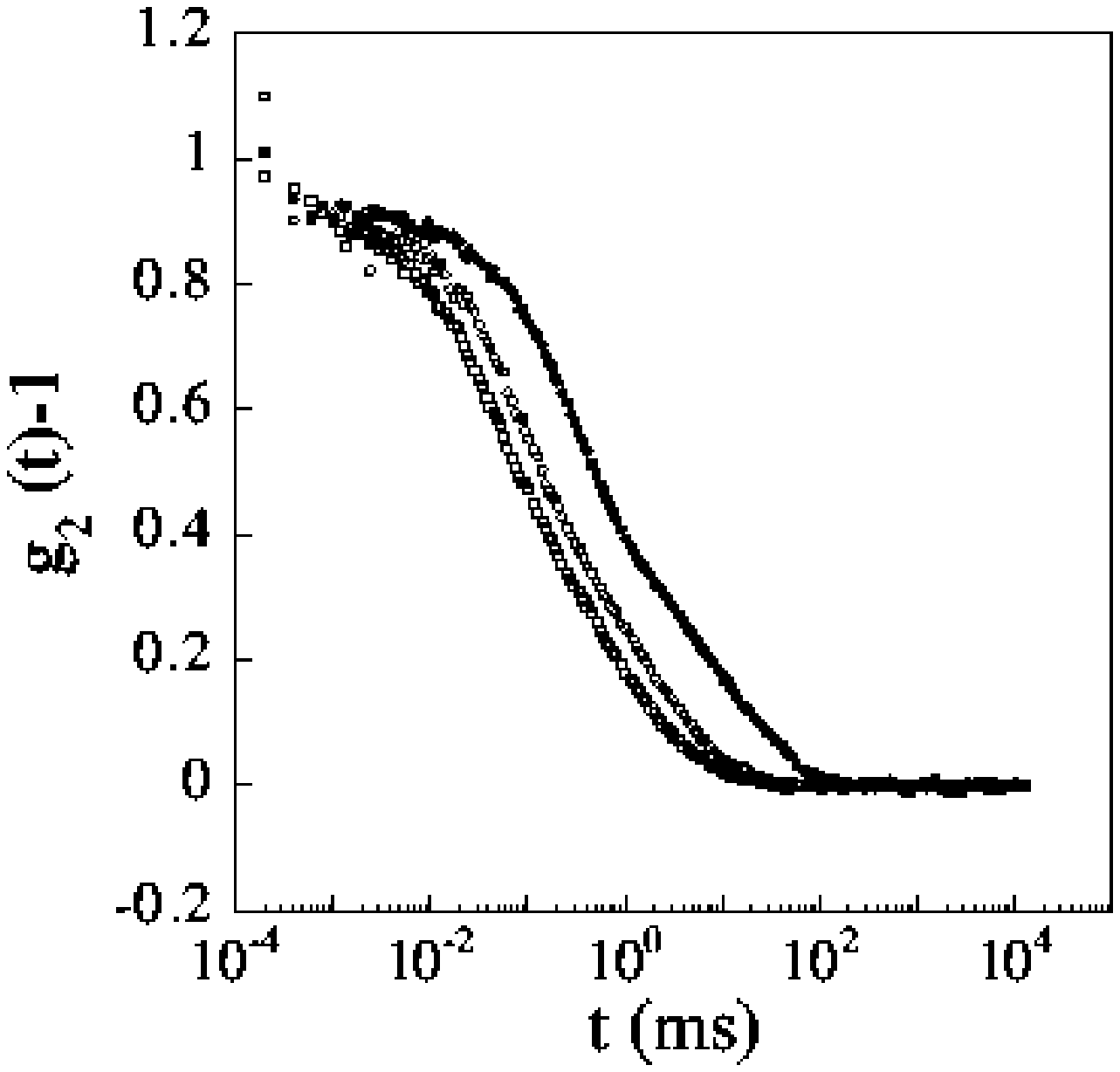}}
\centerline{\epsfxsize=7truecm \epsfbox{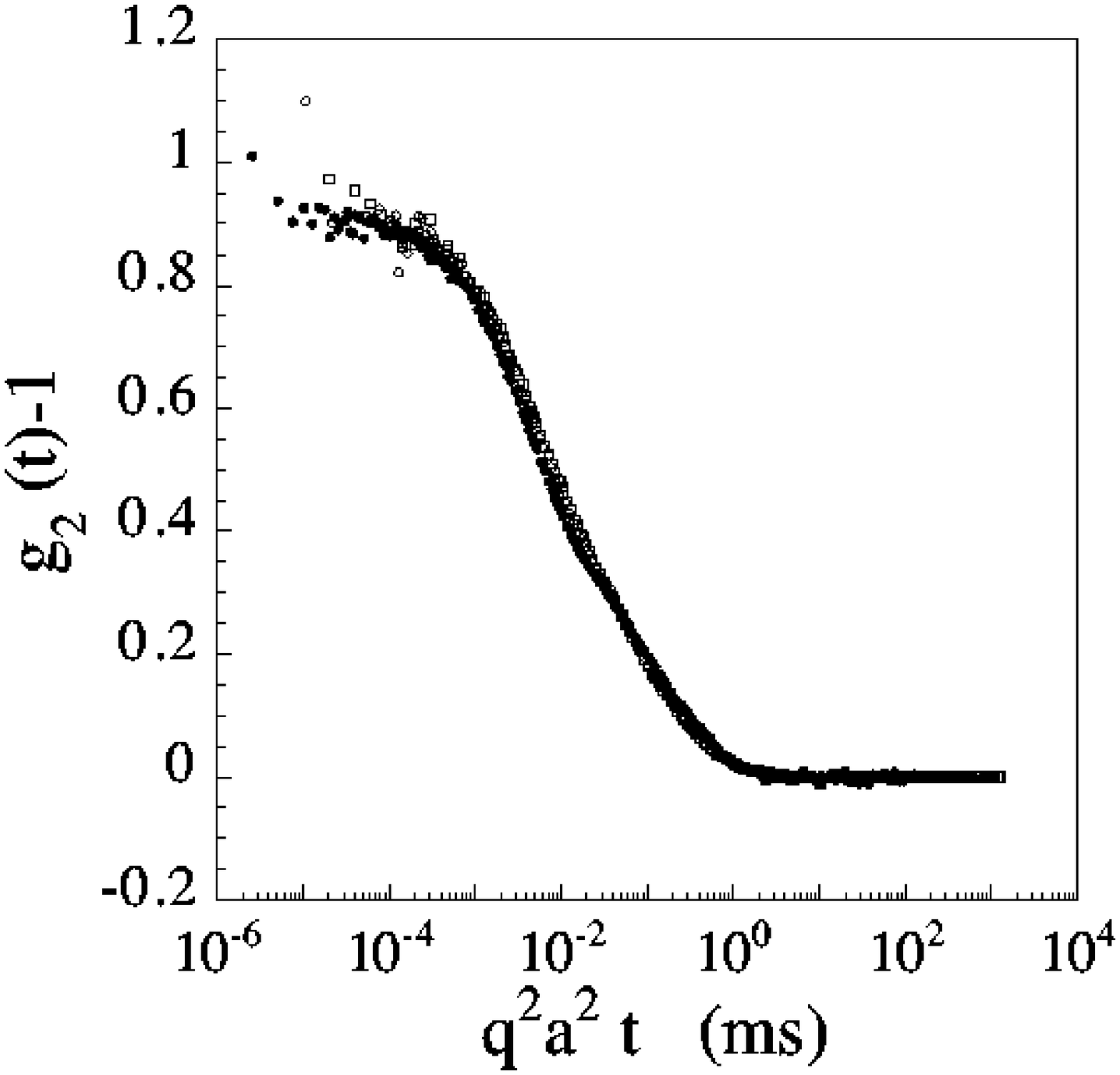}}
\caption{Scaling of the autocorrelation functions taken under
different scattering angles $\theta= 40 ^{\mbox{\tiny{o}}}$ (open
squares), $\theta= 90 ^{\mbox{\tiny{o}}}$ (open circles) and $\theta=
150^{\mbox{\tiny{o}}}$ (filled circles), for an aging time $t_w=300$
min. $a$ is the radius of the particles~\protect\cite{kroon1998}}
\label{scale}
\end{figure}

The observation of the
scaling of $\tau_1$ and $\tau_2$ with $1/q^2$ allows us to rescale the
autocorrelation functions for different scattering angles. Rescaling
the time $t$ as $q^2 a^2 t$, where $a$ is the radius of the Laponite
particles, leads to a collapse of the different correlation functions
(Figure~\ref{scale}). 

\begin{figure}
\centerline{\epsfxsize=8.6truecm \epsfbox{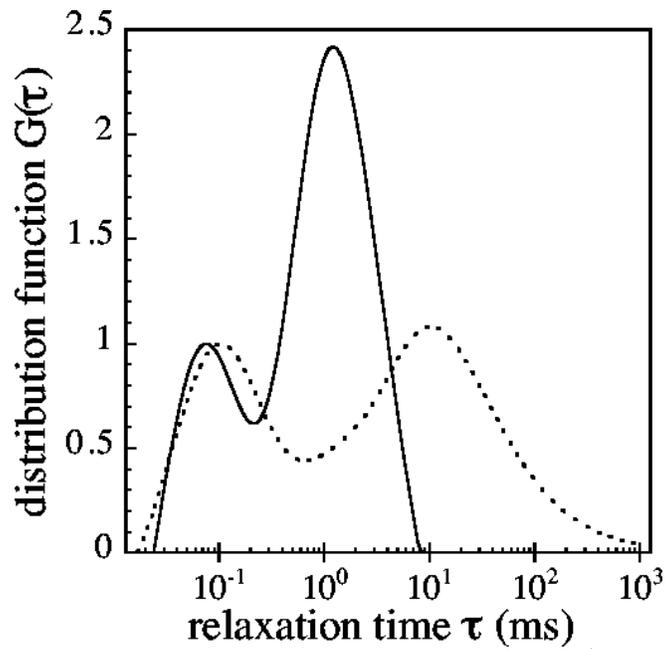}}
\caption{Relaxation time distribution functions corresponding to aging
times $t_w= 0$ (full line) and $t_w=400$ minutes (dashed line) for a $2.5 \%$ wt suspension. The corresponding auto-correlation functions are taken under a scattering angle $\theta=90^{\mbox{\tiny{o}}}$. In order to be able to compare the functions, the peak intensities for the small-scale Brownian motion were normalised to unity.}
\label{decay}
\end{figure}

In order to interpret these results, we determine the distribution of
relaxation times $\tau$ in the system. This we do using a commercially
available constrained regularization method (ALV-NonLin Data
Analysis~\cite{alv}). The method consists in decomposing the intensity
autocorrelation function in exponential modes by directly inverting :
$g_2({\bf q},t)-1=( \int_{\Gamma_{min}}^{\Gamma_{max}} \exp{(- \Gamma t)} G(\Gamma) d \Gamma)^2
$, yielding the distribution function of decay times $G(\tau)$, where $G(\tau)=G(1/\Gamma)$. Figure~\ref{decay} shows the decay time distribution functions $G(\tau)$ corresponding to different aging times $t_w=0$ and $t_w=400$ min. One observes two distinct modes, the fast one corresponding to the first relaxation time $\tau_1$, and the slow one to $\tau_2$. The intensity fast mode has been arbitrarily normalised to unity for the sake of comparison. This fast mode remains almost unchanged as the system ages. The slow relaxation exhibits two different features. First, the maximum of the distribution shifts to larger times, which corresponds to the increase in $\tau_2$ already observed from the direct fit of the correlation function. Second, the distribution of relaxation times $G(\tau)$ becomes wider as the system ages. This corresponds in fact to the decrease in the stretch exponent $\alpha$: the distribution of relaxation times $\tau$ becomes wider and wider. 
These two combined effects make that, very rapidly, the longest
relaxation time in the system exceeds the laboratory time scale of
observation. The direct consequence of this is that, for short aging
times $t_w$, the correlation function decays to zero as shown in
Figure~\ref{inserfit}. For longer aging times ($t_w > 10$ hours for a
$2,5 \%$ wt suspension), the auto-correlation function does not decay
to zero within the observation time scale: the system is no longer ergodic~\cite{bonn-aging}. 

\begin{figure}
\centerline{\epsfxsize=8.6truecm \epsfbox{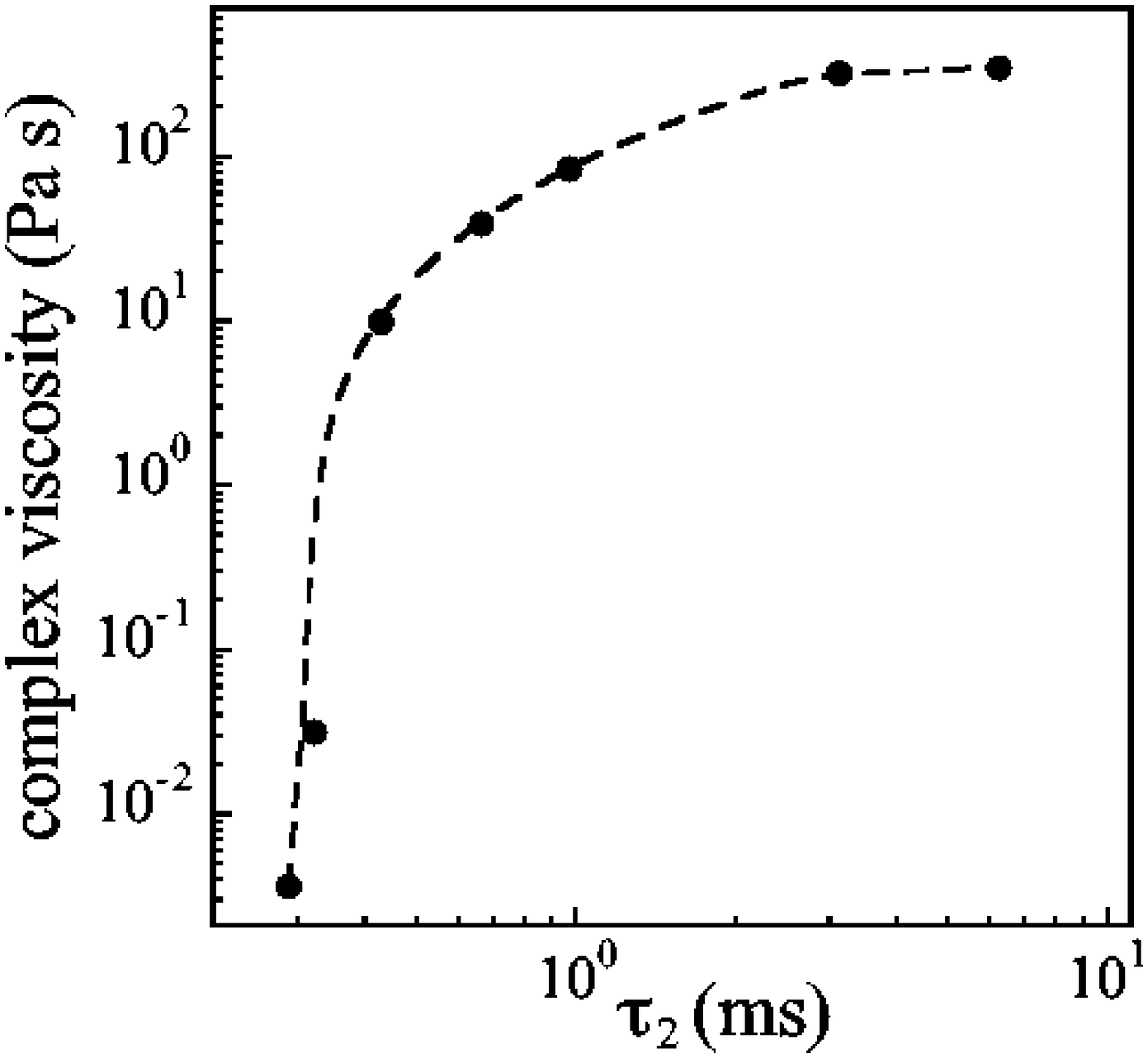}}
\caption{Complex viscosity as a function of $\tau_2$ for a a $2,5 \%$
wt Laponite suspension for a pulsation $\omega=1$ s$^{-1}$.}
\label{viscotau2}
\end{figure}

Experimentally, the dependence of $\tau_2$ and
$\eta^{*}$ on $t_w$ leads to the consequence that the viscosity of the suspension is not
simply proportional to $\tau_2$ as usually assumed in simulations~\cite{onuki,berthier-fluctu,berthier-rheo} (Figure~\ref{viscotau2}).

\section{DISCUSSION AND CONCLUSION}

We propose the following interpretation of our measurements. The diffusion of particles can be described as a cage-diffusion process. The first relaxation $\tau_1$ characterizes the short-time Brownian diffusion of a particle in the suspending liquid. For short times $t$, the particle diffuses freely within a 'cage' formed by the surrounding particles. This diffusive motion consequently does not depend on the aging time. The second relaxation process, occurring for long times $t$, can be interpreted as the escape from the cages. The corresponding characteristic time $\tau_2$ increases rapidly with $t_w$, indicating that it becomes more and more difficult for a particle to 'escape'. This characteristic relaxation time was found also to scale as $1/q^2$, reminescent of diffusive motion. However, this diffusive motion of particles for long times is complex as it is characterized by a broad distribution of relaxation times, leading to the stretched exponential behavior of the correlation function.
From this it follows immediately that for short aging times, the
system is ergodic: the particles reside in the 'cages' formed by
surrounding particles and escape from them after a characteristic time
that depends on the aging time $t_w$. As the system ages, the escape
from the 'cages' becomes slower. The particles are subsequently
constrained by the 'cages', resulting in an ergodicity breaking within
the observation time scale.

The comparison of our experimental results with predictions from
MCT~\cite{bouchaud1996} shows a qualitative agreement with respect to
the slow relaxation (so-called $\alpha$-relaxation) and faster relaxation (so-called $\beta$-relaxation)  features of
the particle dynamics. The non-aging $\beta$-relaxation part would
correspond to the relaxation time $\tau_1$ which is independent of the
aging time $t_w$. Only the slow $\alpha$-relaxation part of the
autocorrelation function is predicted to depend on $t_w$, which is
consistent with the dependence of $\tau_2$ on $t_w$. As in the MCT
scenario, we found that the typical relaxation time $\tau_2$ increases
with increasing $t_w$, which leads to a non-ergodic system when the
largest $\tau_2$ is comparable to the observation time scale. 
In other experiments on aging~\cite{vincent,derec,struik}, the systems
were found to relax on typical
time scales that increase as $t_w^{\mu}$, in
agreement with calculations from mean field
models of aging~\cite{leticia}. When $\mu < 1$, the process of aging is called
sub-aging; this situation is encountered in spin
glasses~\cite{vincent}, in concentrated colloidal
suspension~\cite{derec} or polymer
glasses~\cite{struik}. Recently, diffusing wave spectroscopy (DWS) measurements of the motion
of the tracer particles in glassy Laponite suspensions have also been
performed~\cite{knaebel2000}. The measurements show that the slow
collective relaxation time scales linearly with the waiting time $t_w$
(so that $\mu=1$) called full aging. Interestingly, in our experiments, the typical
relaxation time $\tau_2$ is found to grow exponentially fast with
$t_w$. The log-log inset plot of Figure~\ref{inserttau2} shows that
indeed our data are inconsistent with a power law $\tau_2 \propto
t_w^{\mu}$. In our experiments, the typical decay
time $\tau_2$ of the slow relaxation is of the order of 1 to 10 ms and is consequently seven orders of
magnitude smaller than the aging time $t_w$. In the experiments cited above,
the aging processes were observed in a different regime where the
aging times $t_w$ and the decay times $\tau$ of the slow relaxation are of the same order
of magnitude.

In conclusion, the collective diffusion process of particles can be interpreted as a
cage-diffusion process. For short
aging times, the system is ergodic. The particles reside in the cages
formed by surrounding particles and escape from them after a
characteristic time which is aging time dependent. As the system ages,
the escape from the cages become slower, resulting in a strong
ergodicity breaking. In many systems, such as spin-glasses and
concentrated colloidal supensions, the typical relaxation time is
found to behave as $t_ w^{\mu}$  with $\mu < 1$. Interestingly, in our
experiments, the characteristic time of the slow relaxation process is
found to grow even faster, i.e. exponentially fast with the aging
time. In addition, we find that this dependence on $t_w$ results in a
non-trivial relation between the viscosity and the relaxation time.

\section*{ACKNOWLEDGEMENTS}

LPS de l'ENS is UMR 8550 of the CNRS, associated
with the universities Paris 6 and Paris 7. We thank Leticia
Cugliandolo for helpful discussions. We thank the Ministere de la
Recherche for
financial support (ACI blanche). 

\end{multicols}

\newpage

\newpage

\begin{thebibliography}{99}


\bibitem{bouchaud1996}
Bouchaud, J. P.; Cugliandolo, L.; Kurchan, J.; M\'ezard, M.
\newblock {\em Physica A} {\bf 226}, 243 (1996).

\bibitem{kob1997}
Kob , W.; Barrat, J.
\newblock {\em Phys. Rev. Lett.} {\bf 78}, 4581 (1997). 


\bibitem{nelson}
Nelson, P.; Allen, G. D. (Editors)
\newblock{\it Transport theory and Statistical Physics : Relaxation
kinetics in Supercooled liquids-Mode-coupling theory and its
experimental tests (Marcel Dekker, New-York, 1995)}.

\bibitem{pusey1987}
Pusey, P.N.; Van Megen, W.
\newblock {\em Phys. Rev. Lett.} {\bf 59}, 2083 (1987).

\bibitem{mourchid1995}
Mourchid, A.; Delville, A.; Lambard, J.; Lecolier, E.; Levitz, P.
\newblock {\em Langmuir} {\bf 11}, 1942 (1995).

\bibitem{mourchid1998lang}
Mourchid, A.; Lecolier, E.; Van Damme, H.; Levitz, P.
\newblock {\em Langmuir} {\bf 14}, 4718 (1998).

\bibitem{bonn-aging}
Bonn, D.; Tanaka, H.; Wegdam, G. H.; Kellay, H.; Meunier, J.
\newblock {\em Europhys. Lett.} {\bf 45}, 52 (1999).

\bibitem{kroon1998}
Kroon, M.; Vos, W. L.; Wegdam, G. H. 
\newblock {\em Phys. Rev. E} {\bf 57}, 1962 (1998).

\bibitem{thompson1992}
Thompson, D. W.; Butterworth, J. T.
\newblock {\em J. Colloid
Interface Sci.} {\bf  151}, 236 (1992).

\bibitem{mour1998}
Mourchid, A.;  Levitz, P.
\newblock {\em Phys. Rev.E} {\bf 57}, 4887 (1998).

\bibitem{bonn-filter}
Bonn, D.; Kellay, H.; Tanaka, H.; Wegdam, G. H.; Meunier, J.
\newblock {\em Langmuir} {\bf 15}, 7534 (1999).

\bibitem{pignon1997}
Pignon, F.;  Piau, P.; Magnin, A.
\newblock {\em Phys. Rev. Letter} {\bf 79}, 4689 (1997).


\bibitem{nicolai2000}
Nicolai, T.; Cocard, S.
\newblock {\em Langmuir} {\bf 16}, 8189 (2000).


\bibitem{wigner1938}
Wigner, E.
\newblock {\em Trans. Farad. Soc.} {\bf 34}, 678 (1938).

\bibitem{bosse1998}
Bosse, J.; Wilke, S.D.
\newblock {\em Phys. Rev. Lett.} {\bf 80}, 1260 (1998).

\bibitem{lai1997}
Lai, S.K.; Ma, W.J.; Van Megen, W.; Snook, I.K.
\newblock {\em Phys. Rev. E} {\bf 56}, 766 (1997).


\bibitem{levitz2000}
Levitz, P.; Lecolier, E.; Mourchid, A.; Delville, A.; Lyonnard, S.
\newblock {\em Europhys. Lett.} {\bf 49}, 672 (2000).

\bibitem{alv}
Rainer, P.
\newblock {ALV-NonLin available through ALV-Laser
Vertriebsgesellschaft m.b.H. (1993)}.


\bibitem{berne}
Berne, B.; Pecore, R.
\newblock {\em Dynamic Light Scattering (Wiley, New-York, 1976)}. 

\bibitem{pusey1989}
Pusey, P. N.; Van Megen, W.
\newblock {\em Physica A} {\bf 157 }, 705 (1989).

\bibitem{xue1992}
Xue, J.-Z.; Pine, D. J.; Milner, S. T.; Wu, X.-l; Chaikin, P. M.
\newblock {\em Phys. Rev. A} {\bf 46}, 6550 (1992).

\bibitem{leticia}
Cugliandolo, L.F.;  Kurchan, J.
\newblock {\em Phys. Rev. Lett. } {\bf 71}, 173 (1993). 
\newblock {\em J.  Phys. } {\bf A27 }, 5749 (1994).

\bibitem{vincent}
Vincent, E.; Hamman, J.; Ocio, M.; Bouchaud, J.-P.; Cugliandolo, L.F. 
\newblock {\em Complex behaviour of glassy systems (Springer Verlag
Lect. Notes in Phys., Rubi M. and Perez-Vicente C., Eds., 1997)} {\bf
492}, 184; e-print cond-mat/9607224.

\bibitem{derec}
Derec, C.;  Ajdari, A.;  Ducouret, G.; Lequeux, F.
\newblock {\em C. R. Acad. Sci.}, t1, S\'erie IV, 1115 (2000).

\bibitem{struik}
Struik, L.C.E.
\newblock {\em Elsevier, Houston, 1978}.

\bibitem{bouchaud}
Bouchaud, J.-P
\newblock {\em Soft and Fragile Matter: Nonequilibrium Dynamics,
Metastability and Flow, M. E. Cates and M. R. Evans, Eds.,
IOP Publishing (Bristol and Philadelphia, 2000)}, 285; e-print cond-mat/9910387 and references therein.

\bibitem{knaebel2000}
Knaebel, A.; Bellour, M.; Munch, J.-P.; Viasnoff, V.; Lequeux, F.;
Harden, J.L.   
\newblock {\em Europhys. Lett.}  {\bf 52}, 73 (2000).  

\bibitem{onuki}
Yamamoto, R; Onuki, A.
\newblock {\em Phys. Rev. E } {\bf 58}, 3515 (1998).

\bibitem{berthier-fluctu}
Barrat, J.-L.; Berthier, L.
\newblock {\em Phys. Rev. E} {\bf 63}, 012503 (2001).

\bibitem{berthier-rheo}
Barrat, J.-L.; Berthier, L.; Kurchan, J.
\newblock {\em  Phys. Rev. E} {\bf 61}, 5464 (2000).







\end{thebibliography}
\end{document}